# HIGH STRAIN AND STRAIN-RATE BEHAVIOUR OF PTFE/ALUMINIUM/TUNGSTEN MIXTURES

John Addiss[1], Jing Cai[2], Stephen Walley[1], William Proud[1], Vitali Nesterenko[2, 3]

[1]*Cavendish Laboratory, Madingley Road, Cambridge, CB3 0HE.  UK*
[2]*Materials Science and Engineering Program, University of California San Diego, CA 92093.  USA*
[3]*Department of Mechanical and Aerospace Engineering, University of California San Diego, CA 92093.  USA*

**Abstract.** Conventional drop-weight techniques were modified to accommodate low-amplitude force transducer signals from low-strength, cold isostatically pressed 'heavy' composites of polytetrafluoroethylene, aluminum and tungsten.  The failure strength, strain and the post-critical behavior of failed samples were measured for samples of different porosity and tungsten grain size. Unusual phenomenon of significantly higher strength (55 MPa) of porous composites (density 5.9 g/cm$^3$) with small W particles (<1 μm) in comparison with strength (32 MPa) of dense composites (7.1 g/cm$^3$) with larger W particles (44 μm) at the same volume content of components was observed.  This is attributed to force chains created by a network of small W particles.  Interrupted tests at different levels of strain revealed the mechanisms of fracture under dynamic compression.

## INTRODUCTION

Mixtures containing polytetrafluoroethylene (PTFE) and aluminum (Al) are known to be energetic under dynamic and/or thermal loading [1-4].  They are similar in composition to thermites, a subgroup of the class of pyrotechnics, and are formulated to generate a large quantity of heat during reaction.  The addition of tungsten (W) in this study increases the density and overall strength of the samples.

In this paper the dynamic mechanical properties of cold isostatically pressed composites consisting of PTFE, Al and W powders are investigated using a drop-weight apparatus.  The effects of sample porosity and W particle size on sample strength are determined.

Drop-weight testing allows compressive dynamic loading of samples at a strain rate of approximately 300 s$^{-1}$.  The Cavendish drop-weight apparatus has previously been used to study the deformation of inert materials [5] and the initiation of reaction in thermite mixtures [6].  In this investigation the drop-weight apparatus is modified to accommodate low amplitude force signals from low-strength samples.  The deformation of samples can be interrupted at a chosen strain, thus revealing the mechanisms of fracture under dynamic loading.

## EXPERIMENTAL PROCEDURE

Composites consisting of 77% W (Teledyne, 44 μm or Alfa Aesar, <1 μm), 5.5% Al (Valimet, 2 μm) and 17.5% PTFE (Dupont) by weight were produced by cold isostatic pressing at a range of pressures.  Three different batches were produced giving a variety of porosities and W grain sizes, the details are given in Table 1.  This allowed for comparison of the behavior of samples with the same density but different W particle sizes and with the same size of W particles but different densities.

**Table 1.** Details of the three mixtures which were investigated

| Name | Tungsten particle size/ μm | Pressing Conditions | Average Density/ gcm$^{-3}$ |
|---|---|---|---|
| Porous mixture with fine W | <1 | 350 MPa for 20 min | 6.01 |
| Porous mixture with coarse W | 44 | 20 MPa for 20 min | 6.09 |
| Dense mixture with coarse W | 44 | 350 MPa for 20 min | 7.10 |

The Cavendish drop-weight facility consists of a mass of 6.414 kg which can be dropped from a maximum height of 1.2 m. The falling weight is guided by two external steel guide rods to the impact point where speeds of up to 5 ms$^{-1}$ and strain-rates of a few 100 s$^{-1}$ can be achieved. Force applied to the sample is measured using a transducer consisting of a calibrated Wheatstone bridge arrangement of four strain gauges attached to a maraging steel anvil supporting the sample.

In a conventional drop-weight test the sample is placed between two maraging steel anvils. With this setup drop-weight traces show a considerable degree of scatter and oscillation caused by inevitable vibrations in the device excited by impact. Consequently when testing low strength materials, in which the signal of interest is low relative to the amplitude of the oscillatory noise, it is hard to extract any meaningful information. This problem is illustrated by the drop-weight trace shown in Fig. 1.

In this research conventional tests are modified by the inclusion of a nitrile BS201NI70 o-ring on the upper maraging steel anvil to damp vibrations in the system (Fig. 2). In order to calculate the strain as a function of time it is necessary to take account of the deformation of the o-ring. By carrying out experiments from a variety of heights it was found that the deformation of the o-rings was effectively strain-rate independent. This makes it possible to subtract the height change of the o-ring at a given force from the displacement of the drop-weight at the same force to calculate the strain of the sample as a function of time.

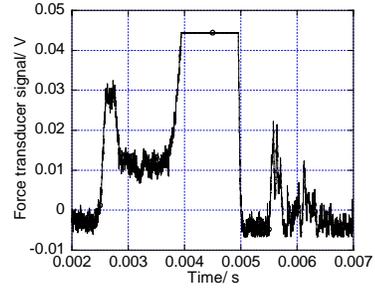

**Figure 2.** Drop-weight trace for a coarse W, porous sample (lowest strength sample) in a "soft" drop-weight test. Oscillations are much reduced relative to the standard method (compare with Fig. 1).

A second consequence of the low strength of soft materials is that they are severely damaged during testing. This makes it difficult to determine the mode of failure. In this research copper rings, with an internal diameter sufficient to allow for the lateral expansion of the sample, were placed around some samples during experimentation in order to interrupt the deformation at a chosen strain. This allowed partially deformed low-strength damaged samples to be recovered after testing.

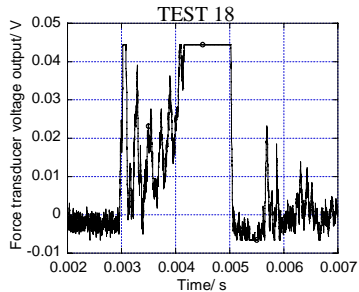

**Figure 1.** Conventional drop-weight trace for a dense mixture with coarse W, the signal of interest is obscured by oscillatory noise.

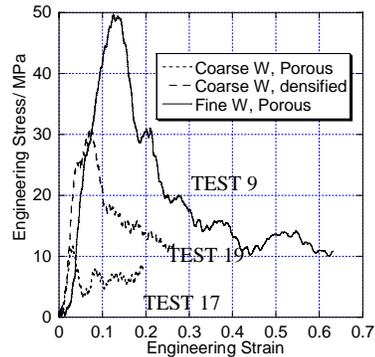

**Figure 3.** Example engineering stress versus true strain plots for the three mixtures described in Table 1.

## RESULTS AND DISCUSSION

Characteristic engineering stress versus engineering strain plots for some of our materials are shown in Fig. 3. There were variations of around 10% in the mechanical properties between samples of the same material, this is to be expected because of their granular nature.

The fine W, porous mixtures were found to have considerably higher failure stresses of around 50 – 55 MPa compared to the coarse W, denser mixtures which failed at stresses of around 30 MPa. This is somewhat different to what might be expected – generally porosity is detrimental to the strength of a material.

A suggested explanation for this phenomena is that force chains are formed in the network of small metal particles leading to a considerable increase in strength relative to the coarse W particle mixtures. Computer simulations have suggested that the strength disparity depends mainly on the distribution of the Al and W particles in the PTFE matrix. The porous matrix in the mixtures with fine W particles does not disrupt the force chains created during the cold isostatic pressing process and during deformation [7]. The result of a simulation illustrating the presence of force chains in a sample at 10% global strain is given in Fig. 4. The stress/force chain formation can be related to ignition sites within the composite energetic material under the compressive load [8, 9].

The dense mixtures with coarse W particles are observed to fail at higher stresses than porous samples with the same W particles (30 MPa compared to 12 MPa). This can be explained by the negative impact porosity has on the strength of a matrix.

Interesting variations were seen in the stress-strain plots for the porous samples with coarse W particles. Some samples exhibited a very low strength and failed at approximately 12 MPa. Other samples exhibited considerably greater strength failing at 35 MPa or above. The stress-strain plots for these stronger samples displayed a long slow rise in stress during the early stages of impact. Examples of both types of plots are shown in Fig. 5.

We attribute the relatively slow rise of stress observed in the plot 2 to a gradual densification of the sample during the initial stage of deformation. This reduction in porosity leads to a considerably increased strength relative to those samples which appear to fail almost immediately upon impact, without the initial densification stage observed in other samples. The failure strengths of the subsequently densified samples are comparable to that of the dense samples with coarse W and are greater than the pressing pressure (20 MPa).

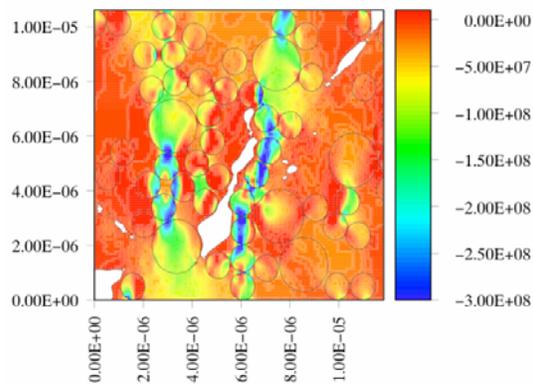

**Figure 4.** The result of a simulation of a drop-weight test (speed =4.43 ms$^{-1}$) on a PTFE-W-Al sample. Force chains are apparent here at 10% global strain. The distance scales are in m [7].

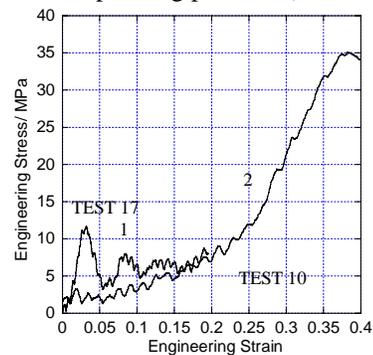

**Figure 5.** Examples of the two types of stress-strain curve obtained for porous samples with coarse W. Curve 1 illustrates practically immediate failure, curve 2 represents densification followed by failure.

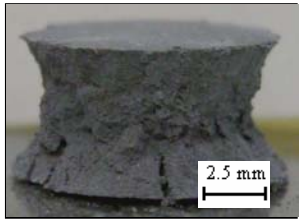

**Figure 6.** Coarse W, porous sample with the deformation interrupted at a strain of approximately 0.1 (Test 73).

Fig. 6 shows a recovered porous sample with coarse W particles where the deformation was interrupted at an engineering strain of 0.1. The stress-strain curve associated with this sample is similar to curve 1 shown in Fig. 5. The sample has failed by shear localization. The relevant meso-scale mechanism of shear localization at low levels of strain is considered in [10], examples of shear localization due to micro-fracture mechanisms leading to reaction in granular materials can be found in [11]. It is uncertain why some samples fail by shearing at low strain while others are densified leading to an increased strength. There is clearly some competition occurring between compaction of the visco-elastic matrix and fracture during the deformation process.

## CONCLUSIONS

The unusual phenomenon of significantly higher strength (55 MPa) of porous composites with small W particles in comparison with strength (32 MPa) of dense composites with larger W particles was observed. This was attributed to the formation of force chains within the network of small tungsten and Al particles, a mechanism which is supported by computer simulation.

Porous samples with coarse W particles were seen to deform by two separate mechanisms. Some failed by shearing at a very low stresses of around 10 MPa. Others appeared to be gradually densified leading to subsequent failure at considerably greater stresses of 35 MPa and at much larger strains.

Polymer o-rings were found to be generally successful at removing oscillatory noise from drop-weight traces making small signals from low-strength materials detectable.

The inclusion of copper rings in some tests in order to interrupt the deformation of samples at a chosen strain was found to be an effective method of investigating the form of deformation occurring.

## ACKNOWLEDGEMENTS

This research was supported by the ONR research award N00014-06-1-0623 and MURI ONR Award N00014-07-1-0740. John W. Addiss is supported by a studentship funded by EPSRC.

## REFERENCES

1. Davis, J. J., Lindfors, A. J., Miller, P. J., Finnegan, S. and Woody, D. L., "Detonation like phenomena in metal-polymer and metal/metal oxide-polymer mixtures", 11[th] Det. Symp. (Int.), 1007-1013, 1998.
2. Holt, W. H., Mock Jr., W., Santiago, F. J., Appl. Phys., 88, 5485, 2000.
3. Ames, R., "Energy release characteristics of impact-initiated energetic materials", Mater. Res. Soc. Symp. Proc., Vol. 896, 123-132, 2006.
4. Denisaev, A. A., Steinberg, A. S. and Berlin, A. A., Doklady Physical Chemistry, Vol. 414, part 2, 139-142, 2007.
5. Swallowe, G. M., Field, J. E., and Horn, L. A., "Measurements of transient high temperatures during the deformation of polymers", J. Mater. Sci. 21, 4089-4096.
6. Walley, S. M., Balzer, J. E., Proud, W. G. and Field, J. E., "Response of thermites to dynamic high pressure and shear", Proc. R. Soc. Lond. A (2000) 456, 1483-1503.
7. Herbold, E.B., Cai, J., Addiss, J. W., Benson, D. J. and Nesterenko, V. F., "The role of porosity in PTFE-W-Al composites under dynamic loading", 17[th] US army symposium on solid mechanics, 2007.
8. Foster J.C. Jr., Greg Glenn, G., and Gunger M., "Meso-scale origins of the low pressure equation of state and high rate mechanical properties of plastic bonded explosives", Shock Comp. of Cond. Matt., 703-706, 1999.
9. Roessig, K.M., "Mesoscale mechanics of plastic bonded explosives", Shock Comp. of Cond. Matt., 973-978, 2001.
10. Dey, T.N., and Johnson, J.N., "Shear band formation in plastic bonded explosive (PBX)", Shock Comp. of Cond. Matt., 285-288, 1998.
11. Nesterenko V.F., Dynamics of heterogeneous materials, Springer-Verlag, NY, 2001.